\documentclass[10pt]{wlscirep}

\usepackage[utf8]{inputenc}
\usepackage[T1]{fontenc}
\usepackage{placeins}
\usepackage{setspace}
\usepackage{adjustbox}
\usepackage{longtable}
\usepackage{pdflscape}
\usepackage{geometry}
\usepackage[table]{xcolor}
\usepackage{longtable}
\definecolor{lightgreen}{HTML}{D5EADF}
\definecolor{lightyellow}{HTML}{FEF1D1}
\definecolor{lightpink}{HTML}{FCE7D7}
\definecolor{headergray}{HTML}{F2F2F2}

\usepackage{lineno}
\usepackage[ruled]{algorithm2e}
\usepackage{soul}

\usepackage{graphicx} 
\usepackage{pdfpages}
\usepackage{tikz}
\usepackage{amsthm}
\usepackage{amsmath}
\usepackage{amssymb}
\usepackage{placeins}

\usepackage{caption}
\usepackage{bm}

\title{Intrinsic Annealing in a Hybrid Memristor-Magnetic Tunnel Junction \\ Ising Machine}

\author[1,+]{Mohammed Akib Iftakher} \author[2,+]{Hugo Levices} \author[1,2]{Kamel-Eddine Harabi} \author[1]{Adrien Renaudineau} \author[2]{Mathieu-Coumba Faye} \author[2]{Corentin Bouchard} \author[3]{Florian Disdier} \author[2]{Bernard Viala} \author[2]{Elisa Vianello} \author[3]{Philippe Talatchian} \author[3]{Kevin Garello} \author[1,*]{Damien Querlioz} \author[2]{Louis Hutin}
\affil[1]{Universit\'e Paris-Saclay, CNRS, Centre de Nanosciences et de Nanotechnologies,  Palaiseau, France}
\affil[2]{Universit\'e Grenoble-Alpes, CEA, LETI,   Grenoble, France}
\affil[3]{Universit\'e Grenoble-Alpes, CEA, CNRS, Grenoble INP, SPINTEC,   Grenoble, France}
\affil[*]{damien.querlioz@universite-paris-saclay.fr}
\affil[+]{These authors contributed equally to this work}

\begin{abstract}
Hardware implementations of the Ising model offer promising solutions to large-scale optimization tasks. In the literature, various  nanodevices have been shown to emulate the spin dynamics for such Ising machines with remarkable effectiveness. Other nanodevices have been shown to implement spin-spin coupling with compact footprint and minimal energy dissipation. However, an ideal Ising machine would associate both types of nanodevices, and they must operate synergistically to support annealing: a progressive reduction of machine stochasticity that allows it to settle to energy minimum. Here, we report an Ising machine that combines two nanotechnologies: memristor crossbar -- storing multi-level couplings -- and stochastic magnetic tunnel junction (SMTJ), acting as thermally driven spins. Because the same read voltage that interrogates the crossbar also biases the SMTJs, increasing this voltage automatically lowers the effective temperature of the machine, providing an intrinsic, nearly circuit-free annealing technique. Operating at zero magnetic field, our prototype consistently reaches the global optimum of a 24-vertex weighted MAX-CUT and a 10-vertex, three-color graph-coloring problem.  Given that both nanotechnologies in our demonstrator are CMOS-integrated, this approach is compatible with advanced 3D integration, offering a scalable pathway toward compact, fast, and energy-efficient large-scale Ising solvers.
\end{abstract}

\begin{document}
\maketitle


\thispagestyle{empty}


\section*{Introduction}

The growing demand for rapid solutions to large, combinatorial optimization tasks -- including resource allocation, scheduling, and AI decision‑making -- has renewed interest in the Ising model\cite{onsager1944crystal}. By mapping a problem onto a network of binary spins that minimize a global Ising Hamiltonian, one can cast many NP‑hard or NP‑complete instances -- graph partitioning, Boolean satisfiability, traveling‑salesman tours -- into a unified framework \cite{Lucas2014}. Physical ‘Ising machines’ accelerate this search by allowing interacting elements to relax naturally toward low-energy configurations, and recent electronic demonstrations based on coupled oscillators\cite{wang2017oscillator,Dutta2021,wang2021solving,ahmed2021probabilistic,moy2021,graber2024integrated,bashar2020experimental,Delacour2023,bazzi2024optimizing,maher2024cmos} or probabilistic bits\cite{borders2019integer,kaiser2022hardware,aadit2022massively,Yin2022TDM,singh2024cmos,si2024energy,Nikhar2024} highlight this momentum.

For practical deployment, an Ising machine must satisfy three intertwined requirements. First, it needs dense, reconfigurable memories that store multi-level spin-spin couplings, yet add negligible static power and area.
Second, it must anneal -- gradually lowering stochasticity during the search -- without resorting to power-hungry supervisory logic. Third,  because routing resources vanish quickly in two dimensions, both the spins and the coupling matrix should ideally reside in CMOS back-end-of-line (BEOL) layers to enable vertical integration.

Here we demonstrate experimentally an Ising machine that satisfies all three requirements by associating two CMOS-integrated nanotechnologies. A hafnium‑oxide memristor crossbar stores the spin-spin couplings, while probabilistic spins are implemented using a perpendicular stochastic magnetic tunnel junction (SMTJ, used as p-bit\cite{camsari2017stochastic, Camsari2019review, Kaiser2021}). 
In our approach, the same read voltage that interrogates the crossbar also biases the SMTJs: lowering this voltage increases spin fluctuations (higher effective temperature), whereas raising it suppresses flips (lower effective temperature). This built-in, device-level mechanism thus provides intrinsic annealing with almost no additional circuitry.
Our prototype solves a 24‑vertex weighted MAX‑CUT and a 10‑vertex, three‑color graph‑coloring problem. All measurements are performed at room temperature and zero external magnetic field. The memristor array and the SMTJ are both integrated above CMOS circuits, on separate dies, providing a feasible route to future three-dimensional integration.

Prior approaches address the problem only partially, relying on nanodevices to implement either spin dynamics or spin-spin couplings. For example, memristor-based Ising machines deliver dense coupling matrices but emulate spins in software or CMOS\cite{CaiNatElec2020,Jiang2022memHNN,jiang2023efficient,kim2024ising,shan2024one}. Conversely, networks of SMTJ p-bits supply intrinsic randomness yet rely on off-chip or on-chip CMOS for spin-spin couplings\cite{borders2019integer,kaiser2022hardware,Yin2022TDM,si2024energy}. The sole hybrid demonstration to date, combining VO\textsubscript{2} oscillators with FeFET couplers\cite{Pantha2024CMOSX}, lacks an annealing path. By integrating memristors with SMTJs and harnessing a voltage-controlled effective temperature, we unify couplings, spins, and annealing in a compact, vertically integrable architecture.

The remainder of this Article describes the device integration and operating principle, presents experimental optimization results, and discusses scalability toward larger problem sizes enabled by three-dimensional integration.

\section*{Results}\label{sec:THR}

\subsection*{Hybrid Ising‑machine concept}

\begin{figure}[h]
    \centering
    \includegraphics[width=1\linewidth]{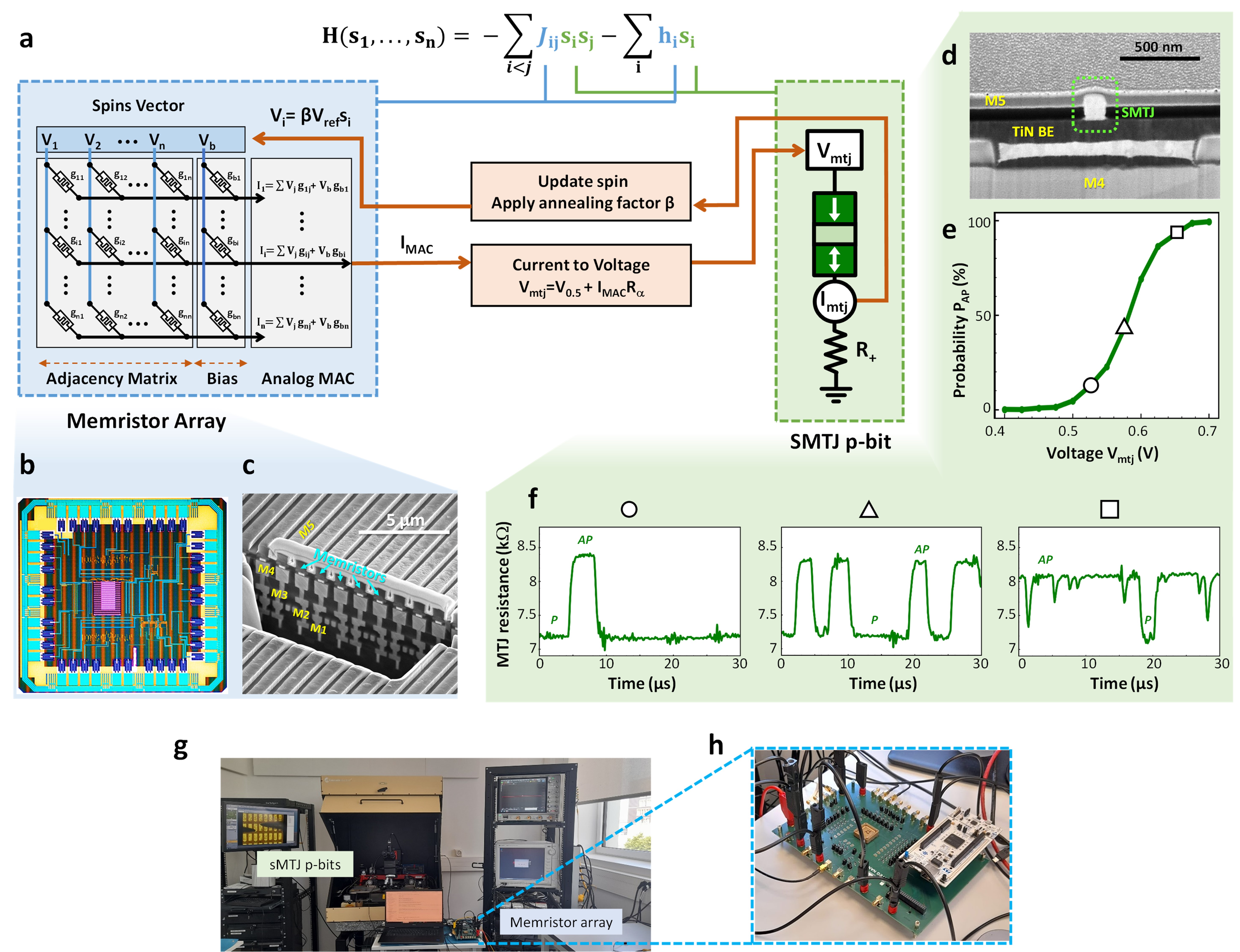}
    \caption{\textbf{Hybrid memristor-SMTJ Ising-machine architecture and device primitives.}  
 \textbf{a}  System-level concept: A hafnium-oxide memristor crossbar computes the weighted sum of neighbouring spins, while a stochastic magnetic tunnel junction (SMTJ) converts that current into a probabilistic spin update. 
 \textbf{b}  Die micrograph of the 32$\times$64 memristor crossbar integrated above CMOS. 
 \textbf{c}  Scanning Electron Microscopy (SEM) image  highlighting the integration of the TiN/HfO$_2$/Ti memristors above the fourth layer of metal interconnects. 
 \textbf{d}  SEM image showing a perpendicular SMTJ integrated in the back-end-of-line (BEOL). 
 \textbf{e} Experimentally determined switching probability of an SMTJ versus applied voltage, showing a sigmoid dependence that implements Gibbs-sampling.  
 \textbf{f} Sample measured time responses of the SMTJ above at $V_{mtj}$ = 525 mV (circle), 575 mV (triangle), 650 mV (square).  \textbf{g} Broader view of the experimental setup.  \textbf{h} The memristor array is mounted on a custom-printed circuit board (PCB) and interfaced with a microcontroller for addressing and communication with the host computer.}
    \label{fig:setup}
\end{figure}

Our objective is to resolve low-energy configurations of the Ising Hamiltonian:
\begin{equation}
  H \;=\; -\sum_{i<j} J_{ij}\,s_i s_j \;-\; \sum_i h_i\,s_i,
  \label{eq:ising_hamiltonian}
\end{equation}
where each spin  $s_i$ takes binary values. The spin-spin coupling, or adjacency matrix \(J\) encodes pairwise interactions, while the vector \(h\) provides on-site biases. The probability of a spin configuration \(\{s\}\) obeys Boltzmann-style statistics,
\begin{equation}
  P\!\bigl(\{s\}\bigr)
  \propto
  e^{-\beta H(\{s\})},
  \qquad
  \beta=\frac{1}{
  T}
  \label{eq:Boltzmann}
\end{equation}
with \(T\) a dimensionless pseudo-temperature.
We generate samples with a canonical, sequential Gibbs sampler. Conditioned on all other spins (which we denote \(s_{\setminus i}\)), the probability for a spin \(s_i\) to be one is
\begin{equation}
  P\!\bigl(s_i = +1 \mid s_{\setminus i}\bigr)
  \;=\;
  \sigma\!\bigl(2\beta f_i\bigr),
  \qquad
  f_i \;=\; \sum_{j} J_{ij}\,s_j \;+\; h_i
  \label{eq:gibbs_sampling}
\end{equation}
where \(\sigma\) is the sigmoid function, and the local effective field $f_i$ sums contributions from neighboring spins, using a multiply-and-accumulate (MAC) operation weighted by \(J_{ij}\) and the on-site bias \(h_i\).
By iterating these updates across all spins, the system randomly explores different configurations in proportion to their Boltzmann weight. 

To avoid premature trapping in local minima, we employ simulated annealing \cite{Kirkpatrick83}: we gradually increase the inverse pseudo-temperature \(\beta\)  over successive updates. Early iterations at low \(\beta\) are highly stochastic, encouraging
broad exploration; as \(\beta\) grows the dynamics become more deterministic, steering the Ising machine toward low-energy states and  -- under a sufficiently slow schedule -- toward global minima of the Hamiltonian.

Fig.~\ref{fig:setup}a overviews our hardware platform, which exploits two complementary nanotechnologies to realize the Gibbs sampling steps in Eq.~(\ref{eq:gibbs_sampling}). First, we compute the local field $f_i$ by a multiply-and-accumulate operation using a memristor crossbar array. Specifically, as shown in Fig.~\ref{fig:setup}b, the memristors (fabricated from hafnium oxide) are integrated in a hybrid CMOS/memristor circuit, with the memristive devices located in the back-end-of-line layers (Fig.~\ref{fig:setup}c, see Methods). When each spin $s_j$ is applied to the corresponding crossbar column, the device conductances 
($g_{ij}=J_{ij}$ and $g_{bi}=h_i$) provide the spin-spin coupling and local biasing, while Kirchhoff’s current summation yields:
\begin{equation}
I_i \;=\; \left( \sum_{j} J_{ij}s_j+h_{i} \right)V_{\text{read}}, 
  \qquad
  V_{\text{read}} \;=\; \beta V_{\text{ref}}
\label{eq:kirchhoff}
\end{equation}
as depicted schematically in Fig.~\ref{fig:setup}d, where $V_{\text{ref}}$ is a reference read voltage. 
This equation is here written in the case of nonnegative coupling $J_{ij}$ and biases $h_i$. (The negative case, which is commonplace in practice, is addressed later in this paper.) This direct electrical summation offers high parallelism and compactness, enabling large-scale networks of spins to be sampled with minimal overhead.

Next, we require a probabilistic element capable of implementing the spin-update rule in Eq.~(\ref{eq:gibbs_sampling}): we need to sample a binary value with a probability given by applying a sigmoid function to the MAC result. 
To achieve this, we leverage stochastic (superparamagnetic) perpendicular magnetic tunnel junction (SMTJ) devices, similar to standard magnetoresistive random-access memory (MRAM) cells but engineered to have extremely low retention times (milliseconds to nanoseconds), making them unstable due to the available thermal noise at room temperature\cite{locatelli2014noise,debashis2016experimental,vodenicarevic2017low,borders2019integer}. As seen in Fig.~\ref{fig:setup}d, these SMTJs are also fabricated in the back-end-of-line (BEOL) of a CMOS process, using an academic process (see Methods). These junctions can fluctuate spontaneously  between parallel and antiparallel magnetic states under thermal agitation. The probability of  each state is modulated by the applied voltage. Figs.~\ref{fig:setup}e,f depict how this probability smoothly transitions from near-zero to near-unity as voltage increases, reproducing the characteristic shape of a sigmoid function. Hence, by feeding the summed current from the memristor crossbar [Eq.~(\ref{eq:kirchhoff})] as the input voltage of an SMTJ-based “p-bit,” we directly implement the probabilistic update prescribed by Eq.~(\ref{eq:gibbs_sampling}).  This data was obtained at zero applied magnetic field,  and we call $V_{0.5}$ the voltage at which the junction has exactly a one half probability of being in each state.

 \FloatBarrier

A key advantage of our approach is that increasing the read voltages $V_{\text{read}}$  naturally increases the effective $\beta$ in Eq.~(\ref{eq:gibbs_sampling}). 
The voltage that we apply  to the SMTJ is $V_{\text{0.5}} + I_{\text{MAC}}R_{\text{$\alpha$}}$, where $R_{\text{$\alpha$}}$ is a conversion constant (see Methods). Lower $V_{\text{read}}$ voltages keep the SMTJ voltage near  $V_{0.5}$, allowing frequent SMTJ switching, and mimicking a high temperature encouraging the system to explore more spin configurations. 
Increasing the read voltages $V_{\text{read}}$ proportionally increases the current $I_{\text{MAC}}$ [Eq.~(\ref{eq:kirchhoff})], bringing the MTJ voltage further from the middle point $V_{\text{0.5}}$. This reduces the likelihood of spontaneous state flips and thereby emulates a lower-temperature regime where spins become more deterministic.  
The Methods section ``Annealing schedule'' 
 shows the mathematical equivalence between our technique and Gibbs sampling with simulated annealing.

This built-in annealing capability eliminates the need for elaborate external schedules, offering a straightforward means to transition from a stochastic state to a stable, low-energy solution. Consequently, the symbiosis between memristor-based MAC operations and intrinsically stochastic SMTJs provides an elegant solution to the annealing challenge that often constrains large-scale hardware Ising machines.


\begin{figure}[h]
    \centering
    \includegraphics[width=1\linewidth]{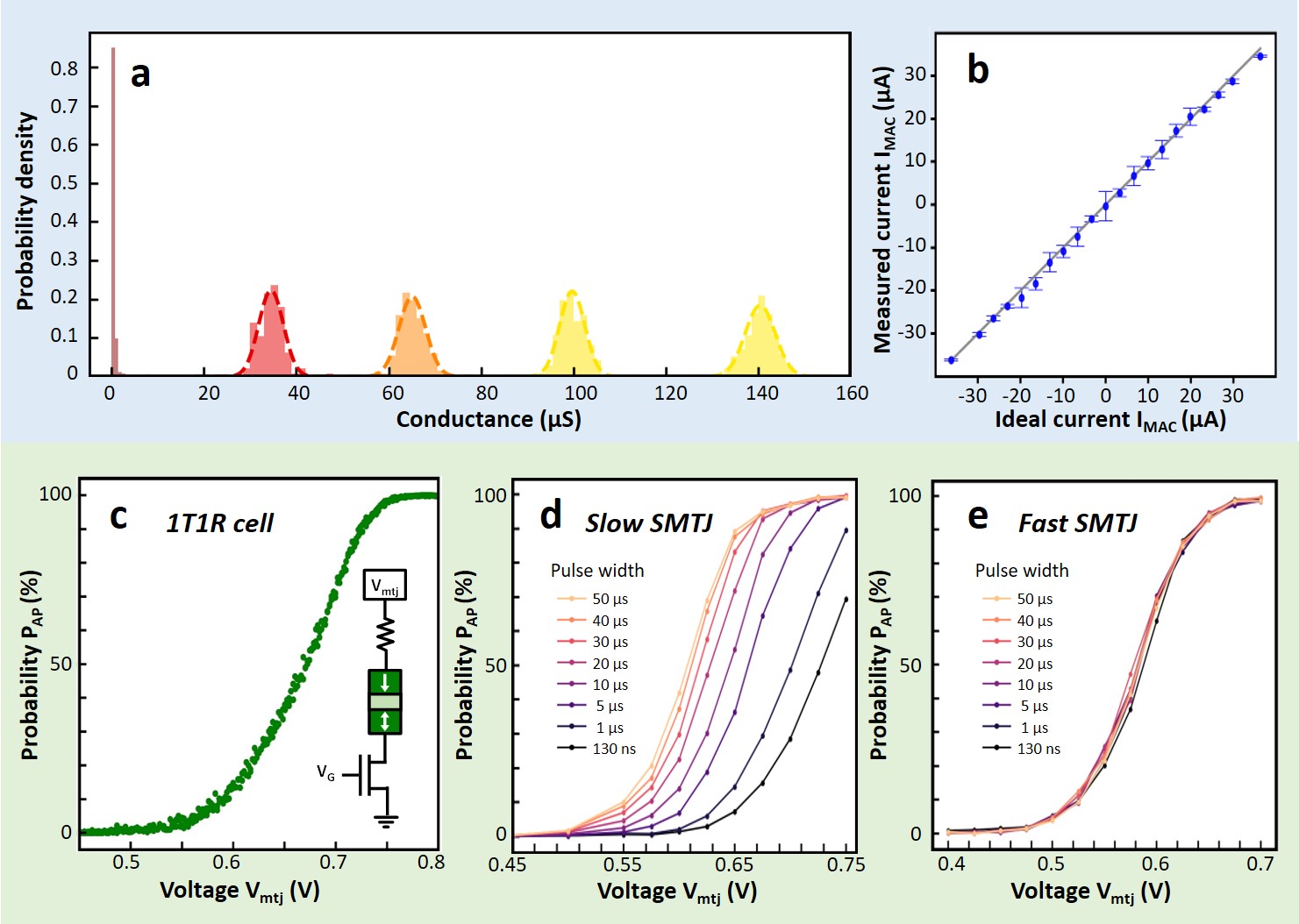}
    \caption{\textbf{Device-level characterisation underpinning reliable, analog coupling and stochastic spins.}  
\textbf{a}  Histograms of programmed memristor state, with target conductance states (0~$\mu$S, 33~$\mu$S, 66~$\mu$S, 99~$\mu$S, and 140~$\mu$S) after programming. 
\textbf{b}  Measured multiply-and-accumulate (MAC) output current versus expected value, validating linear summation of \(J_{ij}\) and \(h_i\) (see Methods). The full line represesents y=x, as a guide for the eyes. Error bars represent one standard deviation of the measured $I_\text{MAC}$. 
\textbf{c} Switching statistics of a single-transistor/single-SMTJ (1T1R) bit-cell, as a function of the voltage across the entire 1T1R structure.
\textbf{d} Pulse-width dependence of the switching probability in a slow SMTJ. The pronounced left-shift of the curves when increasing pulse width, gradually converging to the canonical sigmoidal response ($t_{pulse}$ > 40~µs), highlights the dynamics of thermally-assisted barrier crossing in this lower-speed device. \textbf{e} Same measurements on a faster SMTJ. The traces recorded with pulse widths from 130~ns (instrument-limited minimum) to 50~µs
collapse onto a single sigmoid, signalling
the device’s suitability for high-throughput probabilistic bit generation. 
    }
    \label{fig:more_device}
\end{figure}


\subsection*{Device-level characterization}

\begin{figure}[h]
    \centering
    \includegraphics[width=.8\linewidth]{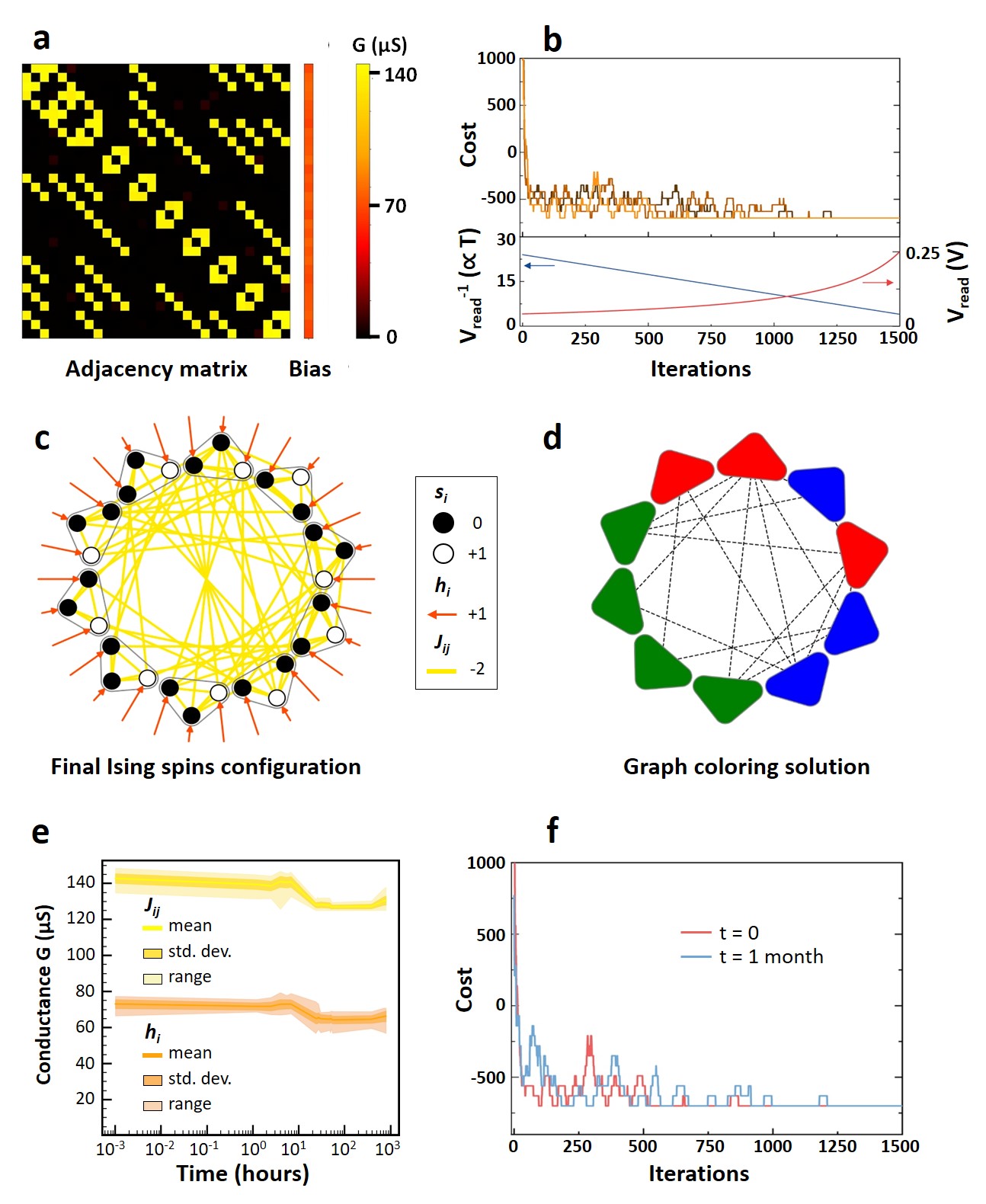}
    \caption{\textbf{Graph-coloring benchmark solved with the same hardware.}  
\textbf{a} One-hot encoding of a ten-node, three-color problem mapped to a sparse 30$\times$30 memristor matrix (140~$\mu$S negative couplings) plus a 70~$\mu$S positive bias column.
\textbf{b}  Linear temperature schedule (1/\(\beta\)) applied by ramping the read voltage and evolution of Ising energy during 1500 sequential spin updates. The three curves are three independent realizations of the experiments using different SMTJs.
\textbf{c,d} Final state of the machine (\textbf{c}; White node: spin of 1, black node, spin of 0), and corresponding color configuration (\textbf{d}) in which no adjacent vertices share a color.
\textbf{e} Mean conductance (solid line), one‑standard‑deviation band (darker shading) and full range (light envelope) of the memristor conductances encoding the couplings $J_{ij}$ (upper traces) and biases $h_i$ (lower traces) measured from $10^{-3}$ h to $10^{3}$ h after programming. Both sets remain within a few~$\mu$S of their initial values over a month, indicating excellent retention of the parameters owing to the RESET-based programming algorithm.
\textbf{f} Cost‑versus‑iteration traces obtained immediately after programming (red) and after one month of room‑temperature storage (blue) for the same graph‑coloring instance. The nearly identical convergence behavior and final cost demonstrate that the small conductance drift observed in (\textbf{e}) does not degrade optimization performance.}
    \label{fig:tasks_results2}
\end{figure}

Fig.~\ref{fig:more_device}a provides further characterization of the hafnium oxide memristors and their programming methodology. To achieve stable, finely tuned conductance levels for accurate MAC operations (Fig.~\ref{fig:more_device}b), we employ a program-and-verify procedure, which iteratively checks and refines each device’s conductance  (see Methods). As shown by the histograms, this approach yields narrow distributions of programmed conductance values across the array. Notably, unlike the more common SET-based programming approach -- which relies on controlling the compliance current -- we instead rely on the RESET mechanism. This method proves superior for long-term device stability, as confirmed by experiments shown later in Fig.~\ref{fig:tasks_results2}.

The SMTJs are fabricated through an academic back-end-of-line process on a commercially produced CMOS integrated circuit (see Methods). Fig.~\ref{fig:more_device}c confirms that the expected sigmoidal relationship between switching probability and voltage is preserved when the device is embedded in an on‑chip one‑transistor/one‑resistor (1T1R) bit‑cell. To the best of our knowledge, these results constitute the first demonstration of stochastic magnetic tunnel junctions monolithically integrated onto a commercial CMOS chip.

Because the academic process does not yet guarantee perfect free‑layer thickness uniformity, the SMTJs exhibit significant die-to-die variability. Figs.~\ref{fig:more_device}d,e present data from devices that switch at different speeds. These measurements, performed at zero magnetic field, reveal the minimum pulse width that still produces a sigmoid-like charasteristic for each SMTJ. For every tested pulse duration, we applied 1,000 identical read pulses and recorded the fraction of antiparallel (AP) outcomes. In Fig.~\ref{fig:more_device}d, pulses shorter than $\sim 2\,\mu\text{s}$ deliver  spin‑transfer torque  during a time insufficient  to reach the upper tail of the sigmoid curve.  
A $5\,\mu\text{s}$ pulse already spans almost the full $0\text{-}100\,\%$ probability range, making it adequate for Ising‑machine updates, while extending the pulse beyond $40\,\mu\text{s}$ produces no further change—behavior consistent with the modified Néel-Brown thermal‑activation model~\cite{brown1963thermal,rippard2011thermal,li2004thermally}. 

By contrast, the fast SMTJ in Fig.~\ref{fig:more_device}e retains an identical sigmoid down to the instrument‑limited minimum pulse width of 130~ns, implying that perpendicular SMTJs of this type can support update rates well above 10~MHz. Even faster, nanosecond‑scale switching at low voltage has been reported previously in non‑CMOS processes~\cite{hayakawa2021nanosecond,safranski2021demonstration,soumah2024nanosecond,schnitzspan2023nanosecond}.

\FloatBarrier

\begin{figure}[h]
    \centering
    \includegraphics[width=1\linewidth]{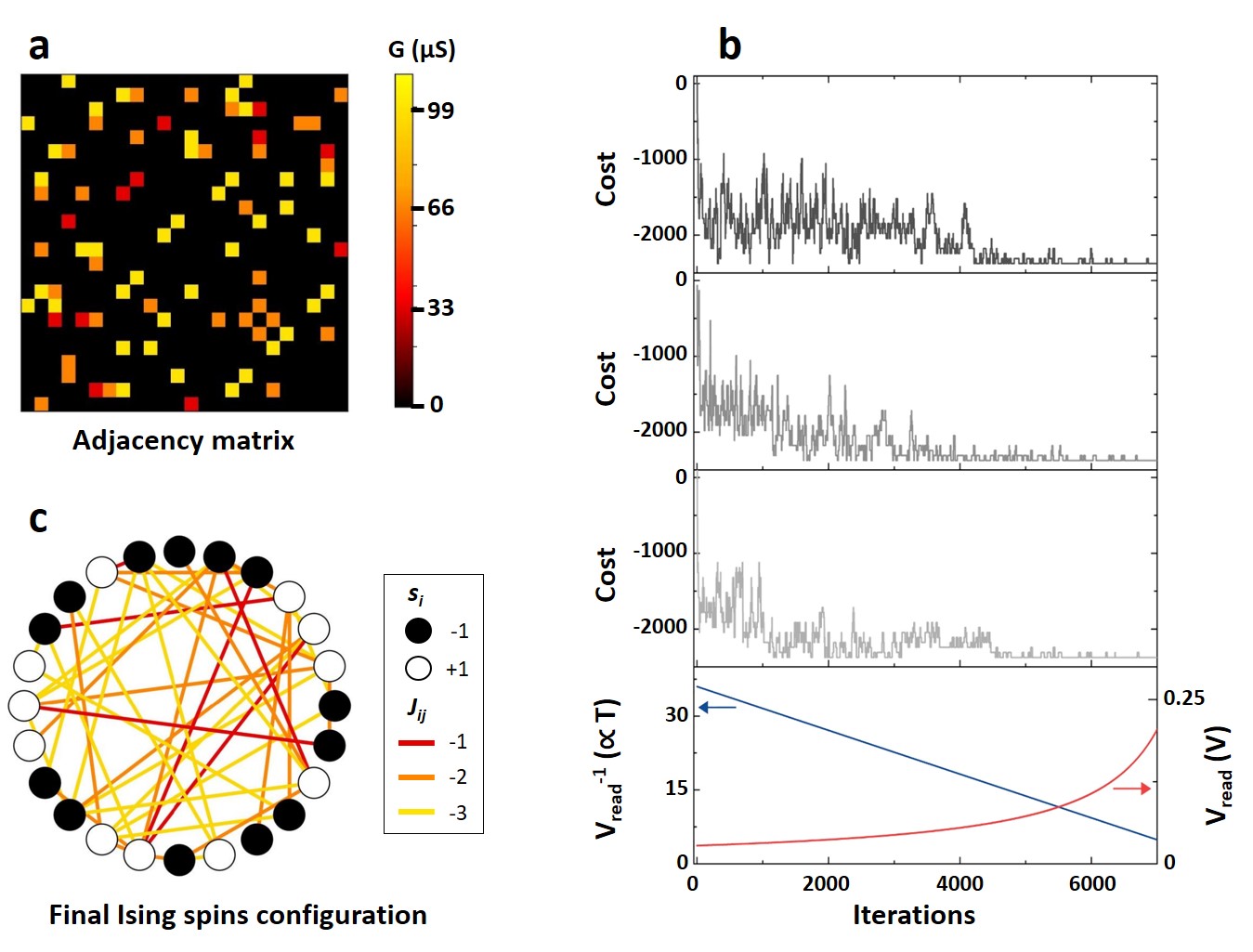}
    \caption{\textbf{Hardware solution of a 24-node weighted MAX-CUT instance.}  
\textbf{a} Conductance map programmed into the memristor array, implementing a three-level \(J_{ij}\) matrix (33~µS, 66~µS, 99~µS), and \(h_i=0\)). 
\textbf{b} Linear temperature schedule (1/\(\beta\)) applied by ramping the read voltage $V_{\text{read}}$ and evolution of Ising energy during 7200 sequential spin updates; the system escapes local minima early and converges to the optimum cut predicted by a software preliminary study. The three curves are three independent realizations of the experiments using different SMTJs.
\textbf{c} Experimentally obtained final state solving the MAX-CUT task (white node: +1 spin, black node: -1 spin).
    }
    \label{fig:tasks_results1}
\end{figure}

\vspace{1cm}

\subsection*{Solving a weighted MAX‑CUT instance}

With the device primitives validated, we next ask whether our two-nanotechnologies Ising machine  can solve an optimization problem.
We first evaluated it on the weighted maximum cut (MAX-CUT) problem, a classic benchmark in the field and an NP-hard task with applications ranging from circuit partitioning to combinatorial optimization. As illustrated in Fig.~\ref{fig:tasks_results1}, the objective is to divide a graph’s nodes into two sets that maximize the total weight of edges cut by the partition. 

Our experiment (shown in Figs.~\ref{fig:setup}g,h) is performed using the CMOS-integrated memristor array shown in Figs.~\ref{fig:setup}b,c. In our 24$\times$24 demonstration, the coupling matrix terms $J_{ij}$  must support multiple discrete levels (see Methods), which we implement directly via the analog conductance states of the memristors (in this instance, all bias terms $h_i=0$). In this task, the spins assume values +1 and -1, which translate into voltages $+V_{\text{read}}$ and $-V_{\text{read}}$ on the memristor array (see Methods).
Fig.~\ref{fig:tasks_results1}a shows the measured adjacency matrix programmed into our memristor array, demonstrating precise setting of conductance values.

To solve MAX-CUT, we evaluate spin values one-at-a-time to carry out the Gibbs sampling steps. We compute the value of the field $f_i$ of one spin using the memristor array. The result is then  directly applied to a unique  SMTJ (Fig.~\ref{fig:setup}d) to sample a corresponding spin value along Eq.~(\ref{eq:gibbs_sampling}). The process is then repeated, each spin one-at-a-time.  We progressively increased $V_{read}$ throughout the process, thereby lowering the effective temperature (Fig.~\ref{fig:tasks_results1}b, see Methods). This annealing schedule enabled the Ising system to escape local minima in early iterations and converge to a stable, low-energy configuration after around 7,000 spin updates (Fig.~\ref{fig:tasks_results1}c). Across multiple experimental trials (shown in Fig.~\ref{fig:tasks_results1}b), the machine reliably found the correct MAX-CUT solution,  with some variation in the exact number of iterations required. 
(We confirmed using a software solver that the final cost obtained experimentally is the task optimum and that the final state is a correct solution.) 
Each of the trial shown in Fig.~\ref{fig:tasks_results1}b was performed using a different SMTJ. This reproducibility underscores the system’s robustness and confirms that the hybrid memristor-SMTJ architecture can effectively handle multi-level spin couplings, a key requirement for solving complex, weighted Ising problems.

\subsection*{Solving a graph coloring instance with bias terms}

To further showcase the versatility of our hardware Ising machine, we tested it on the graph coloring problem -- another NP-hard task with broad relevance in scheduling, register allocation, and frequency assignment. In this problem, each node must be assigned one of several colors, with the constraint that adjacent nodes cannot share the same color. We targeted a 10-node graph colored with three colors, a setting that can be achieved using purely binary $J_{ij}$ values and non-zero biases $h_i$ (see Methods). 
Each color assignment is mapped to a spin subsystem, resulting in a 30$\times$30 system of coupled spins whose adjacency matrix and biases are programmed onto the memristor array (Fig.~\ref{fig:tasks_results2}a). 
In this task, the spins assume values one and zero. As the coupling values $J_{ij}$ are nonpositive, we represent the spins using voltages $-V_{\text{read}}$ and 0 onto the memristor array. Conversely, as the bias is positive, a voltage of $V_{\text{read}}$ is applied onto the bias column of the memristor array.
Note that this technique of choosing the sign of the read voltage to represent the sign of Hamiltonian coefficients only works if all the coefficients in a column have the same sign, which is the case here. For tasks where it is not, it would possible to use differential memristor structures to implement signed Hamiltonian coefficients, as is usually done to implement signed synapses in memristor-based neural networks \cite{prezioso2015training,wan2022compute}.

Following the same annealing schedule concept (Fig.~\ref{fig:tasks_results2}b), the machine consistently converged to a valid coloring in all experimental trials using three different SMTJs (Fig.~\ref{fig:tasks_results2}b,c,d).
We also confirmed using a software solver that the final cost is the task optimum. 
Convergence  occurred more quickly than for the MAX-CUT task, reflecting the simpler binary coupling structure in the graph coloring formulation.

We exploited the graph-coloring benchmark to test the long-term retention of the programmed spin-spin couplings and biases. Fig.~\ref{fig:tasks_results2}e tracks the mean conductance of the coupling (140~$\mu$S) and bias (70~$\mu$S) memristors over a one-month period while the packaged chip was stored at room temperature in ambient air. The drift remains below 
15~$\mu$S for both device populations, confirming the excellent stability of the RESET-programmed states. Without re-programming the array, we reran the graph-coloring experiment thirty days later; Fig.~\ref{fig:tasks_results2}f shows that the solver still converges to the optimal coloring. (Because the SMTJ probe station was unavailable at that time, the spin-update sigmoid was emulated in software in the thirty-days experiment; only the memristor crossbar was operated in hardware.) This result demonstrates that the modest conductance drift does not impair computational accuracy.

\FloatBarrier

\section*{Discussion}

We have shown experimentally that a hybrid memristor/SMTJ platform can solve non-trivial Ising problems while integrating three key functions directly in hardware: multi‑level spin-spin couplings, stochastic spins, and voltage‑controlled annealing.  Both a 24‑vertex weighted MAX‑CUT and a 30‑spin graph‑coloring instance were solved at room temperature and zero magnetic field.

Table~\ref{tab:tableau} places our results in the context of previous nanoelectronic Ising machines.  To the best of our knowledge,  only one earlier demonstrator combines a nanotechnology for spins and a nanotechnology for spin-spin couplings \cite{Pantha2024CMOSX}; however, because its spins are encoded in the discretized phase of oscillators, implementing simulated annealing is a considerable challenge and was not demonstrated.  All other prototypes fall back on CMOS for either the spins or the coupling matrix, limiting scalability in latency, area, or energy -- points we examine below.

Our present prototype uses Gibbs sampling, with spins sampled one at a time. Still, it is fully compatible with parallel operations as the memristor array naturally computes the field of several spins simultaneously. Achieving such parallelism simply demands a set of physical p-bits that can switch concurrently. SMTJs meet this requirement because a single junction, plus a minimal selector, forms a p‑bit; equivalent quality CMOS implementations of p-bits can consume thousands of transistors \cite{singh2024cmos}.  Therefore, our approach would operate ideally  with the multiple parallel extensions of Gibbs sampling that have been introduced in recent years to accelerate convergence.  Chromatic Gibbs sampling updates all non‑interacting spins simultaneously by exploiting graph coloring \cite{aadit2022massively,Nikhar2024}; replica‑based tempering exchanges configurations among copies held at staggered effective temperatures \cite{ISAKOV2015265,lee2025noise}; and fully asynchronous dynamics allow each spin to flip whenever its local conditions are met \cite{borders2019integer,AaditNano,si2024energy}.  
Device‑level refinements such as variability‑aware biasing and in‑situ learning of couplings \cite{kaiser2022hardware} could further boost scalability.

Such a parallel version of our approach has the potential to be very fast.
Our work uses, to the best of our knowledge, the first CMOS-integrated SMTJs. Our fastest junctions sample p-bits in 130~ns. By contrast, state‑of‑the‑art, non-CMOS-integrated SMTJs switch in nanoseconds at femtojoule energies \cite{hayakawa2021nanosecond,soumah2024nanosecond,safranski2021demonstration,schnitzspan2023nanosecond}, two to three orders of magnitude better than all‑CMOS p‑bits \cite{singh2024cmos}. The memristor crossbars, which we use in this work to implement spin-spin couplings, also have the potential for nanosecond operation,  because they do not need an ADC at the output: the MAC current feeds into an SMTJ. This means that our approach can, in principle, sample spins in parallel, at nanosecond speeds. Previous studies have shown that this capability can reduce time to solution by several orders of magnitude compared with conventional electronic implementations \cite{aadit2022massively,singh2024cmos}.

Memristor crossbars are the gold standard for energy-efficient multiply‑and‑accumulate operations, and, until now, have been mostly used for low‑energy in‑memory computing of neural networks \cite{prezioso2015training,le202364,ambrogio2023analog,wan2022compute,aguirre2024hardware}.  For Ising workloads the benefits are even larger than for neural networks, as we do not need an ADC. Additionally, the annealing schedule needs only two global DACs ($V_{\text{read}}$ and $-V_{\text{read}}$), not one high‑resolution DAC per input (see Supplementary Fig.~\ref{fig:parallel_updates}). Finally, when the target Ising graph is sparse -- as in our benchmarks, where $85.4\,\%$ of MAX‑CUT and $82\,\%$ of graph‑coloring couplings were zero—most memristors can remain in a high‑resistance, near‑zero‑power state, making the architecture energetically advantageous. 

In our current prototype, the memristor arrays and SMTJ-based stochastic units are separated by centimeter-scale PCB traces and laboratory-bench cabling. Achieving the targeted nanosecond-level coupling therefore requires tighter integration eliminating all off-board interconnects.  Because both devices are already fabricated in the CMOS BEOL, they are compatible with multiple advanced integration techniques. A pragmatic roadmap for scaling our machine begins with 2.5-D integration: relocating the two dies onto a silicon interposer removes board-level parasitics while leveraging existing advanced-packaging infrastructure. The next milestone is 3-D hybrid bonding, in which prefabricated wafers are joined face-to-face through copper-copper/oxide-oxide bonds, delivering sub-micron pitches without the cumulative thermal budget of sequential BEOL steps \cite{ParkECTC, Gorchichko}. The ultimate goal is monolithic 3-D integration, where layer-by-layer fabrication places both devices in the same BEOL stack. Demonstrations of multi-tier memristor arrays connected by sub-100 nm vias confirm the manufacturability of such stacks \cite{Chakrabarti2017,Li2023,liu2024edge}, and technology roadmaps identify monolithic 3-D as pivotal for future high-density, high-bandwidth compute-in-memory and neuromorphic architectures \cite{WongIEDM2024,vianello2024scaling}. Across this progression, interconnect lengths shrink from millimeters (2.5-D) to micrometers (hybrid bonding) and ultimately to sub-micrometers (monolithic 3-D), bringing memory-access latencies in line with the intrinsic nanosecond dynamics of SMTJs.


A further consideration for scalability is the resolution of memristor conductance values. The RESET- programmed HfO$_x$ memristors used here met the accuracy demands of our benchmarks, but larger or more intricate problems may require finer conductance granularity.  Higher‑precision analog memristors \cite{rao2023thousands,park2024multi,stecconi2022filamentary} or multi‑cell techniques already employed in neural networks accelerators \cite{le202364,ambrogio2023analog} can supply that resolution.  Such devices would also enable in‑situ training and dynamic retuning of couplings \cite{laydevant2024training}, broadening the scope of hardware Ising machines.

By unifying memory, compute and randomness in a vertically integrable stack, hybrid memristor-SMTJ hardware offers a credible path toward compact, highly energy efficient optimizers capable of tackling real‑time scheduling, embedded AI and scientific computing tasks that are currently impractical on von Neumann machines.  This work paves the way for a fully integrated prototype that would shift the state of the art in hardware accelerators for combinatorial optimization.


\FloatBarrier

\section*{Acknowledgements}
This work benefited from  France 2030 government grants managed by the French National Research Agency (ANR-22-PEEL-0009, ANR-22-PEEL-0010 and ANR-22-PEEL-0015), and from Horizon Europe grant 101182279 in the frame of Chips JU FAMES Pilot Line. The authors would like to thank L.~Herrera Diez and F.~Mizrahi for discussion and invaluable feedback.  Parts of this manuscript were revised with the assistance of a large language model (OpenAI ChatGPT).

\section*{Author contributions statement}
M.A.I. designed the electrical test setup under the supervision  of K.E.H. and with the assistance of A.R. M.A.I. performed the electrical experimental work with assistance of H.L and K.E.H. H.L. performed the experimental preparation work and performed the data analysis with the assistance of M.A.I. K.E.H designed the memristor array chip with the assistance of M.C.F. E.V. led the fabrication of the memristor arrays. K.G. and B.V. oversaw the fabrication of the magnetic tunnel junctions, which was performed by C.B. and F.D.
P.T. performed the initial testing of the magnetic tunnel junctions. L.H. and D.Q. directed the work and wrote the initial version of the manuscript.
All authors discussed the results and reviewed the manuscript.

\section*{Competing interests}
The authors declare no competing interests.

\section*{Data availability}
The data measured in this study are available from the corresponding author upon request.

\section*{Code availability} 
The software programs used for modeling Ising machines are available from the corresponding author upon request.


\FloatBarrier

\section*{Methods}

\subsection*{Fabrication of the memristor arrays}

The memristor arrays used in our experiments were fabricated by following a three-phase process flow, building on techniques reported in previous integrated-circuit studies \cite{harabi2023memristor, jebali2024powering, bonnet2023bringing}. In the first phase, the complementary metal-oxide-semiconductor (CMOS) portion was produced at a commercial foundry employing a 130-nm low-power process with four metal interconnect layers. During the second phase, the hafnium oxide (HfO$_x$) memristors were integrated between the fourth and fifth metal layers. Each memristor features a stacked structure of titanium nitride (TiN), HfO$_x$, titanium (Ti), and a top TiN layer, with the 10-nm HfO$_x$  active layer deposited by atomic layer deposition (ALD). The Ti interface layer was also set to 10 nm, while each memristor device was patterned to a 300-nm diameter. Finally, a fifth metal interconnect layer was deposited above the memristors, aligned over vias to ensure reliable electrical contact.

In the third phase, the completed wafers underwent packaging in J-leaded Ceramic Chip Carrier (JLCC) modules, handled by a commercial vendor. This packaging step ensures robust protection and facilitates efficient testing, eliminating the need for specialized probe stations or probe cards in subsequent measurements.

\subsection*{Design and programming of the memristor arrays}

Our memristor platform is designed for unconventional, memristor-based computations. It comprises a 32$\times$64 two-transistor-one-resistor (2T1R) array, where each cell includes two transistors and one memristor. Although the array is capable of both vertical and horizontal transistor addressing, here we operate it in a 1T1R configuration by employing only the horizontal addressing. The on-chip peripheral circuitry supports multiple operational modes. (1) A device-characterization mode provides the infrastructure to develop forming and programming techniques, which we used in this study. (2) An analog computing mode enables the memristors to perform multiply-and-accumulate (MAC) operations by multiplying an input voltage vector with memristor conductance values. In our work, this capability underpins the Ising Hamiltonian operations, representing spins by the former and encoding the adjacency matrix through the latter. Additional modes of operation, not relevant to the present experiments, are also available but were not utilized here.

Forming and programming pulses were applied by a Keysight B1530A waveform generator/fast measurement unit. We formed each memristor once, applying twenty-microsecond voltage pulses of gradually increasing amplitude until the device reached a resistance of approximately 10~k$\Omega$. For subsequent programming, we adopted a program-and-verify scheme, issuing successive ten-microsecond RESET pulses (voltages ranging between 2.0~V and 2.5~V) until achieving the target conductance. The verification step used the Keysight B1530A in fast IV mode, with a multi-second delay after each pulse to ensure stable settling of the device’s resistive state, following guidelines in \cite{esmanhotto2022experimental}. Unlike \cite{esmanhotto2022experimental}, however, our approach relies on RESET pulses (rather than SET pulses) to achieve more stable and reproducible conductance levels.

\subsection*{Fabrication of the stochastic magnetic tunnel junctions}
As with the memristor arrays, the fabrication of the SMTJ chip used in our experiments begins with a CMOS process, followed by the deposition of four metal interconnect layers. This front-end fabrication is performed at a commercial foundry using a 130-nm process. After forming a TiN bottom electrode with low surface roughness, the 200-mm wafers are downsized to 50~mm for subsequent processing. The magnetic tunnel junction stack is deposited in the following sequence: Ta(7) / W(1.5) / CoFeB(1.1) / MgO(RA = 6.7~$\Omega.\mu m^2$) / CoFeB(1.1) / W(0.3) /  [Co(0.6)/Ta(0.2)/Pt(1.1)]$_3$ / Co(0.6) / Ru(0.9) / Co(0.6) / [Co(0.5)/Pt(0.25)]$_6$ / Ru(5), where subscripted brackets indicate the number of repetitions in each multilayer, and numbers in parentheses denote  nominal thicknesses in nanometers (RA corresponds to the resistance-area product, which strongly depends on the thickness of the insulating MgO layer). The first CoFeB layer is the free layer; the second one is the reference layer, which is strongly pinned by the neighboring synthetic antiferromagnetic (SAF)  structure. The SAF is composed of two Co/Pt muli-layer blocks antiferromagnetically coupled through an RKKY interaction, provided by a thin Ru interface layer separating the two blocks.

Following stack deposition, the SMTJs are patterned by electron beam lithography with a nominal diameter of 60~nm. From electrical characterization, we estimate an electrical diameter eCD = 36~nm. Each junction is locally encapsulated with a spin-on polymer, which is patterned and then recessed to expose the top of the conductive hard mask. 
A fifth interconnect layer is subsequently defined using Cr/Al 200-nm metallization and a lift-off process.

After patterning, the effective perpendicular anisotropy of the free layer is sufficiently small that its magnetization can switch in a stochastic manner because of the available thermal noise when a low current is applied, at room temperature and at zero magnetic field. The effective perpendicular anisotropy remains positive, allowing to maintain the magnetization out-of-plane.

\subsection*{Measurements of stochastic magnetic tunnel junctions}

\paragraph{Initial device characterization.}
To extract the voltage-dependent switching characteristics of the studied magnetic tunnel junctions, we applied voltage pulses   across an electrical branch composed of the SMTJ in series with a static resistor $R_{+}$ of 1~k$\Omega$. To evaluate the resistance of the SMTJ we sample the current $I_{\text{mtj}}$ flowing in the branch at given times using a Keysight B1530 waveform generator/fast measurement unit. 
The investigated SMTJs exhibit two metastable resistance states: a low-resistance parallel state (\(R_{\text{P}}\)), and a high-resistance antiparallel state (\(R_{\text{AP}}\)). We determine the state of the junction by comparing its measured resistance to the mean point between \(R_{\text{P}}\) and \(R_{\text{AP}}\), i.e.,  $\overline{R_{\text{mtj}}} = \dfrac{R_{\text{AP}} + R_{\text{P}}}{2}$. 

The waveforms presented in Fig.~\ref{fig:setup}f show measurements where the state of the junction is sampled every 130~ns. To obtain the sigmoids-like curves presented in Fig.~\ref{fig:setup}e and in Fig.~\ref{fig:more_device}, for each input voltage $V_{\text{mtj}}$, we applied 1000 voltage pulses of duration 50~$\mu$s to the device, with current measurements at times 130~ns, 1~$\mu$s, 5~$\mu$s, 10~$\mu$s, 20~$\mu$s, 30~$\mu$s, 40~$\mu$s, 50~$\mu$s. We plot the proportion of these measurements where the junction was in the AP state. (Fig.~\ref{fig:setup}e shows this proportion at time 50~$\mu$s, while Fig.~\ref{fig:more_device}d,e show it at all measured times).

In the resulting curves, we see that as the voltage pulse amplitude $V_{\text{mtj}}$ is swept from lower to higher  values, the extracted probability $P_{\text{AP}}$ increases from zero to one following a sigmoid-like pattern.
We fitted this dependence of $P_{\text{AP}}$ with respect to $V_{\text{mtj}}$ by a sigmoid function:
\begin{equation}
P_{\text{AP}}(V_{\text{mtj}})
=
\frac{1}{1 + e^{-K\,(V_{\text{mtj}} - V_{\text{0.5}})}}\,.
\label{eq:sigmoidmtj}
\end{equation}
Here, $K$ and \(V_{\text{0.5}}\) are constant parameters of the fit which respectively control the slope (positive) of the sigmoid function, and the voltage at which $P_{\text{AP}}$ is exactly 0.5. This empirical relationship forms the basis for our device model for simulation and informs subsequent inference-phase parameters.

\paragraph{When performing optimization tasks.}
When the machine operates in inference mode, the memristor crossbar outputs a MAC current \(I_{\text{MAC}}\). This current is converted to a voltage pulse across the SMTJ branch using
\begin{equation}
V_{\text{mtj}} = V_{\text{0.5}} + I_{\text{MAC}}R_{\text{$\alpha$}},
\label{eq:defvmtj}
\end{equation}
where $R_{\text{$\alpha$}}$ is a conversion constant. In our experiments, we use a value 
$R_{\text{$\alpha$}}=2\bigl(\overline{ R_{\text{mtj}} } + R_{+}\bigr)$.
To avoid junction damage, the applied voltage $V_{\text{mtj}}$ never exceeds a compliance voltage set to be below the SMTJ's breakdown voltage.
For each 50-\(\mu\)s reading pulse, we acquire a single measurement of the SMTJ’s resistance, which is then converted into a binary state 
by comparing it to the threshold resistance level $\overline{ R_{\text{mtj}} }$.

This direct, single-point readout closely mimics how the on-chip comparator would operate in a standalone implementation. It also guards against artificially smoothing the intrinsic stochastic behavior of the SMTJ, thus preserving an accurate measure of the device’s random telegraph switching and ensuring authentic probabilistic updates in the Ising machine.

\subsection*{Experimental setup to solve MAX-CUT and graph coloring tasks}
The memristor-array integrated circuit was mounted on a custom-printed circuit board (PCB) and interfaced with an STMicroelectronics STM32F746ZGT6 microcontroller, which provided digital I/O signals for array addressing and communication with the host computer. Two Keysight B1530A waveform generator/fast measurement channels supplied programming pulses for the memristors, as well as the $V_{\text{read}}$  signal during analog MAC operations. A Keithley 2230G triple-channel power supply provided stable bias voltages for all on-board components.

For the stochastic magnetic tunnel junction  measurements, individual SMTJ dies were probe-tested using a 25-probe card managed by two Keysight B1530A channels, supported by a Keysight MSOS204A oscilloscope and a controlling computer. A 1~k$\Omega$ external resistor $R_+$was added in series to permit real-time observation of the SMTJ’s binary fluctuations on the oscilloscope. The B1530A applied input signals to the SMTJ and recorded the resulting currents, while the control computer processed these data in real time to determine the SMTJ state or, inversely, to generate the input signals for the memristor array.

All equipment was orchestrated through a Python-based environment in a Jupyter notebook, which ran the MAX-CUT and graph-coloring experiments. The notebook issued instructions to the microcontroller and to the B1530A units in a loop, alternating between memristor programming and read operations, SMTJ state measurement, and adjustments to 
$V_{\text{read}}$  in accordance with the annealing schedule. This setup enabled synchronized, closed-loop control of both the memristor crossbar and the SMTJ p-bits for seamless hardware Ising machine operation.


\subsection*{Mapping tasks to the Ising model}

Our benchmark graphs were synthetically generated with sizes and weight distributions selected to match the throughput of our hardware experiment. For each instance, an exhaustive enumeration on a classical workstation confirmed that the solution obtained by our machine experimentally was the global optimum.

\paragraph{Weighted MAX-CUT.}
For a graph \(G=(\mathcal V,\mathcal E)\) with vertex set \(\mathcal V=\{1,\dots,N\}\) and weighted edges
\(W_{uv}>0\;(u,v)\in\mathcal E\),
MAX-CUT maximizes 
\(\tfrac12\sum_{(u,v)\in\mathcal E}W_{uv}\bigl(1-s_u s_v\bigr)\)
with binary spins \(s_u\in\{-1,+1\}\).
Discarding the constant term yields the Ising Hamiltonian  
\begin{equation}
  H
  = A\sum_{(u,v)\in\mathcal E}W_{uv}\,s_u s_v ,
  \label{eq:maxcut_H}
\end{equation}
where the positive scale factor \(A\) is arbitrary and can be used to match device ranges.
Identifying the generic Ising form  
\(
H=-\sum_{i<j}J_{ij} s_i s_j-\sum_i h_i s_i,
\)
we directly obtain  
\begin{equation}
  J_{uv}=-A\,W_{uv},\qquad h_i=0 
\end{equation}

Each non-zero \(J_{uv}\) is realized by programming the corresponding memristor to one of three conductance levels (33, 66, 99~$\mu$S),
proportional to the allowed edge weights.
Because the spin-spin couplings \(J_{ij}\) are nonpositive, we apply the read voltage with inverted polarity:
a logical spin \(s=+1\) is driven by \(-V_{\mathrm{read}}\) and \(s=-1\) by \(+V_{\mathrm{read}}\).
The MAX-CUT formulation contains no on-site fields, hence the bias column is left unprogrammed \((h_i=0)\).

\subsection*{Mapping graph coloring to the Ising Hamiltonian}

We consider an undirected graph \(G=(\mathcal V,\mathcal E)\) with
\(|\mathcal V|=N\) vertices and a fixed palette of \(C\) colors.
A one-hot binary variable  
\(s_{v,k}\in\{0,1\}\) is introduced for each vertex \(v\in\mathcal V\) and color \(k\in\{1,\dots,C\}\):
\(s_{v,k}=1\) if and only if vertex \(v\) is painted with color \(k\).
The problem is cast in quadratic unconstrained binary optimization
(QUBO) form following Ref.~\cite{Lucas2014}:

\begin{equation}
H
=A\sum_{v\in\mathcal V}\Bigl(1-\sum_{k=1}^{C}s_{v,k}\Bigr)^{2}
\;+\;
A\sum_{(u,v)\in\mathcal E}\sum_{k=1}^{C}s_{u,k}\,s_{v,k},
\label{eq:HCOL_QUBO}
\end{equation}
where the first term enforces the one-hot condition (exactly one color per vertex) and the second term penalizes equal colors on adjacent vertices.  Expanding the square in~\eqref{eq:HCOL_QUBO} and using \(s_{v,k}^{2}=s_{v,k}\) yields

\begin{equation}
H
=2A\sum_{v\in\mathcal V}\!\sum_{k<c}s_{v,k}s_{v,c}
+2A\sum_{(u,v)\in\mathcal E}\sum_{k=1}^{C}s_{u,k}s_{v,k}
-A\sum_{v\in\mathcal V}\sum_{k=1}^{C}s_{v,k}+ \text{const}.
\label{eq:HCOL_expanded}
\end{equation}

To match the standard Ising notation  
\(H=-\sum_{i<j}J_{ij}s_i s_j-\sum_i h_i s_i\)
we index the binary variables as  
\(i = C\,(v-1)+k\) and \(j = C\,(u-1)+c\),
so that the coefficient mapping is

\begin{equation}
\begin{aligned}
J_{ij}&=-2A
&&\text{if } 
\bigl(v=u\text{ and }k\neq c\bigr)
\;\;\text{(one-hot)}\\
J_{ij}&=-2A
&&\text{if } 
\bigl(u,v\bigr)\in\mathcal E\text{ and }k=c
\;\;\text{(adjacency)}\\
h_i&=A & &\text{for all } i .
\end{aligned}
\label{eq:J_h_colour}
\end{equation}

All non-zero couplings therefore share the same magnitude \(|J_{ij}|=2A\) and are negative, favoring opposite spin values.
In hardware, we realize these entries with identical conductances
(\(2A=140\,\text{\textmu S}\) in our prototype) and drive them with
\(-V_{\mathrm{read}}\).
The positive fields \(h_i=A\) are implemented with a dedicated bias column
(\(A=70\,\text{\textmu S}\)) accessed by \(+V_{\mathrm{read}}\).
The resulting \(J\) matrix is sparse -- every row contains at most \(C-1\) intra-vertex links and \(|\mathcal N(v)|\) inter-vertex links.

 \subsection*{Annealing schedule}

We implement annealing by raising the voltage applied to the memristor crossbar in proportion to the inverse pseudo-temperature parameter \(\beta\). Specifically, we define
\begin{equation}
V_{\text{read}} = \beta\,V_{\text{ref}}.
\end{equation}
We exploit the fact that our SMTJs exhibit a voltage-dependent probability [Eq.~(\ref{eq:sigmoidmtj})]. If we reinject the voltage $V_{\text{mtj}}$ [Eq.~(\ref{eq:defvmtj})] in Eq.~(\ref{eq:sigmoidmtj}), and the result of the MAC operation [Eq.~(\ref{eq:kirchhoff})],  we find 
\begin{equation}
P_{\text{$AP$}} 
= 
\frac{1}{1 + e^{- \beta K V_{\text{ref}}R_\alpha  \left( \sum_{j} J_{ij}s_j+h_{i} \right)   }}.
\end{equation}
This result maps to the equation of Gibbs sampling [Eq.~(\ref{eq:gibbs_sampling})], with $\beta$ playing the role of a pseudo-temperature. 

In practice, we use a linear schedule that decreases \(T\) (or equivalently increases \(\beta\)) from a high initial value (corresponding to a read voltage $V_{\text{read}}$ in the 30~mV to 40~mV range), which promotes frequent spin flips, to a low final value (corresponding to a read voltage $V_{\text{read}}$ in the 250~mV range) , where spins become nearly deterministic. The temperature is changed every 50 iterations. The exact annealing schedules are shown in Figs.~\ref{fig:tasks_results1}b and~\ref{fig:tasks_results2}b. We use a different schedule for both tasks, as our benchmark graph coloring task can converge faster than our benchmark weighted MAX-CUT task. The annealing schedules were chosen based on prior simulations of the Ising machine. 
The solver’s pseudo-temperature is governed by \(V_{\mathrm{read}}\) and its conversion into
MAC current through the memristor crossbar.
At the beginning of the annealing process, \(V_{\mathrm{read}}\) assumes low values, which might be difficult to generate precisely on-chip, and where memristor conductance can be partly nonlinear. However, precise control of these low voltages is not critical:
Annealing schedules are tuned empirically to perform robustly across families of problem
instances rather than to hit an exact temperature at each iteration, and the smallest
\(V_{\mathrm{read}}\) values coincide with the high-temperature phase, where stochastic
fluctuations dominate spin updates and damp the effect of small voltage errors.


\bibliography{references}  


\clearpage

\begin{landscape}
\scriptsize
\begin{longtable}{|>{\centering\arraybackslash\bfseries}m{3cm}|*{6}{>{\centering\arraybackslash}m{2.5cm}|}}
\hline
\rowcolor{headergray}
Feature & \cellcolor{lightgreen}\textbf{This work} & Cai et al. (2020) \cite{CaiNatElec2020} & Jiang et al. (2023) \cite{jiang2023efficient} & Singh et al. (2024) \cite{singh2024cmos} & Kim et al. (2024) \cite{kim2024ising} & Shan et al. (2024) \cite{shan2024one} \\
\hline
Spins &
\cellcolor{lightgreen}\textbf{Nano (SMTJ)} &
\cellcolor{lightpink}Software &
\cellcolor{lightpink}Software &
\cellcolor{lightgreen}Nano (SMTJ-driven clocking of LFSR) &
\cellcolor{lightyellow}Thresholding unit (comparator) &
\cellcolor{lightyellow}Thresholding unit (comparator) \\
\hline
Coupling &
\cellcolor{lightgreen}\textbf{Nano (Memristors)} &
\cellcolor{lightgreen}Nano (Memristors) &
\cellcolor{lightgreen}Nano (Memristors) &
\cellcolor{lightyellow}FPGA &
\cellcolor{lightgreen}Nano (Memristors) &
\cellcolor{lightgreen}Nano (Memristors) \\
\hline
Biases &
\cellcolor{lightgreen}\textbf{Nano (Memristors)} &
\cellcolor{lightpink}No &
\cellcolor{lightpink}No &
\cellcolor{lightyellow}FPGA &
\cellcolor{lightpink}No &
\cellcolor{lightpink}No \\
\hline
Annealing Method &
\cellcolor{lightgreen}\textbf{Intrinsic: $I_{\text{MAC}}$ modulation through $V_{\text{read}}$} &
\cellcolor{lightpink}Variable noise level introduced in thresholding function (software) &
\cellcolor{lightpink}Gradual non-convexity annealing (software) &
\cellcolor{lightyellow} $\beta$ adjustable digitally &
\cellcolor{lightyellow}Bit Line read noise at low $V_{\text{read}}$ or low G to emulate high temperature &
\cellcolor{lightyellow}Decaying feedback conductance on the thresholding comparator (hardwired RC delay) \\
\hline
Problems Solved (Size) &
\parbox{2.3cm}{\textbf{\centering Weighted MAX-CUT\\(24, sparse)\\Graph Coloring\\(10 nodes, 3 colors)}} &
\parbox{2.3cm}{\centering MAX-CUT\\(32, 50\% density)} &
\parbox{2.3cm}{\centering MAX-CUT\\(64, 50\% density)\\Weighted MAX-CUT (64)\\TSP (9 cities)} &
\parbox{2.3cm}{\centering Invertible\\Full Adder\\Deep Boltzmann\\machine learning} &
\parbox{2.3cm}{\centering MAX-CUT\\(32, 50\% density)} &
\parbox{2.3cm}{\centering MAX-CUT\\(96, sparse)} \\
\hline
\end{longtable}
\setcounter{table}{0}
\scriptsize
\begin{longtable}{|>{\centering\arraybackslash\bfseries}m{3cm}|*{6}{>{\centering\arraybackslash}m{2.5cm}|}}
\hline
\rowcolor{headergray}
Feature & Yin et al. (2024) \cite{yin2024ferroelectric} & Borders et al. (2019) \cite{borders2019integer} & Kaiser et al. (2022) \cite{kaiser2022hardware} & Yin et al. (2022) \cite{Yin2022TDM} & Si et al. (2024) \cite{si2024energy} & Pantha et al. (2024) \cite{Pantha2024CMOSX} \\
\hline
Spins &
\cellcolor{lightpink}Software &
\cellcolor{lightgreen}Nano (SMTJ) &
\cellcolor{lightgreen}Nano (SMTJ) &
\cellcolor{lightgreen}Nano (SMTJ) &
\cellcolor{lightgreen}Nano (SMTJ) &
\cellcolor{lightgreen}Nano (VO$_2$-based oscillator) \\
\hline
Coupling &
\cellcolor{lightgreen}Nano (FeFET) &
\cellcolor{lightyellow}MCU+DAC &
\cellcolor{lightyellow}MCU+DAC &
\cellcolor{lightyellow}MCU+DAC &
\cellcolor{lightyellow}MCU+DAC &
\cellcolor{lightgreen}Nano (BEOL FeFET) \\
\hline
Biases &
\cellcolor{lightpink}No &
\cellcolor{lightyellow}MCU+DAC &
\cellcolor{lightyellow}MCU+DAC &
\cellcolor{lightyellow}MCU+DAC &
\cellcolor{lightyellow}MCU+DAC &
\cellcolor{lightpink}No \\
\hline
Annealing Method &
\cellcolor{lightpink}Software &
\cellcolor{lightpink}No &
\cellcolor{lightpink}No &
\cellcolor{lightpink}No &
\cellcolor{lightgreen}MCU corrects update current through SMTJ vs. temp. &
\cellcolor{lightpink}No \\
\hline
Problems Solved (Size) &
\parbox{2.3cm}{\centering Graph Coloring\\(7 nodes, 3 colors)} &
\parbox{2.3cm}{\centering Factoring 945\\Invertible AND Gate} &
\parbox{2.3cm}{\centering Invertible\\Full Adder} &
\parbox{2.3cm}{\centering Factoring 945} &
\parbox{2.3cm}{\centering TSP\\(9 cities)} &
\parbox{2.3cm}{\centering MAX-CUT (10)\\Spin Glass (10)\\Weighted MAX-CUT (10)} \\
\hline
\caption{\textbf{Hardware implementations of nanotechnology-based Ising machines.} Green: implemented with nanotechnology. Yellow: implemented with digital electronics. Red: simulated in software or not implemented. SMTJ: stochastic magnetic tunnel junction, MCU: microcontroller unit, DAC: digital-to-analog converter, FPGA: Field-Programmable Gate Array, TSP: traveling salesman problem. 
\label{tab:tableau}}
\end{longtable}
\end{landscape}

\clearpage

\renewcommand{\thesection}{S\arabic{section}}
\renewcommand{\theequation}{S\arabic{equation}}
\renewcommand{\thefigure}{S\arabic{figure}}
\renewcommand{\thetable}{S\arabic{table}}

\setcounter{section}{0}
\setcounter{equation}{0}
\setcounter{figure}{0}
\setcounter{table}{0}

\section*{Supplementary Material}
\addcontentsline{toc}{section}{Supplementary Material}

\begin{figure}[h]
    \centering
    \includegraphics[width=\linewidth]{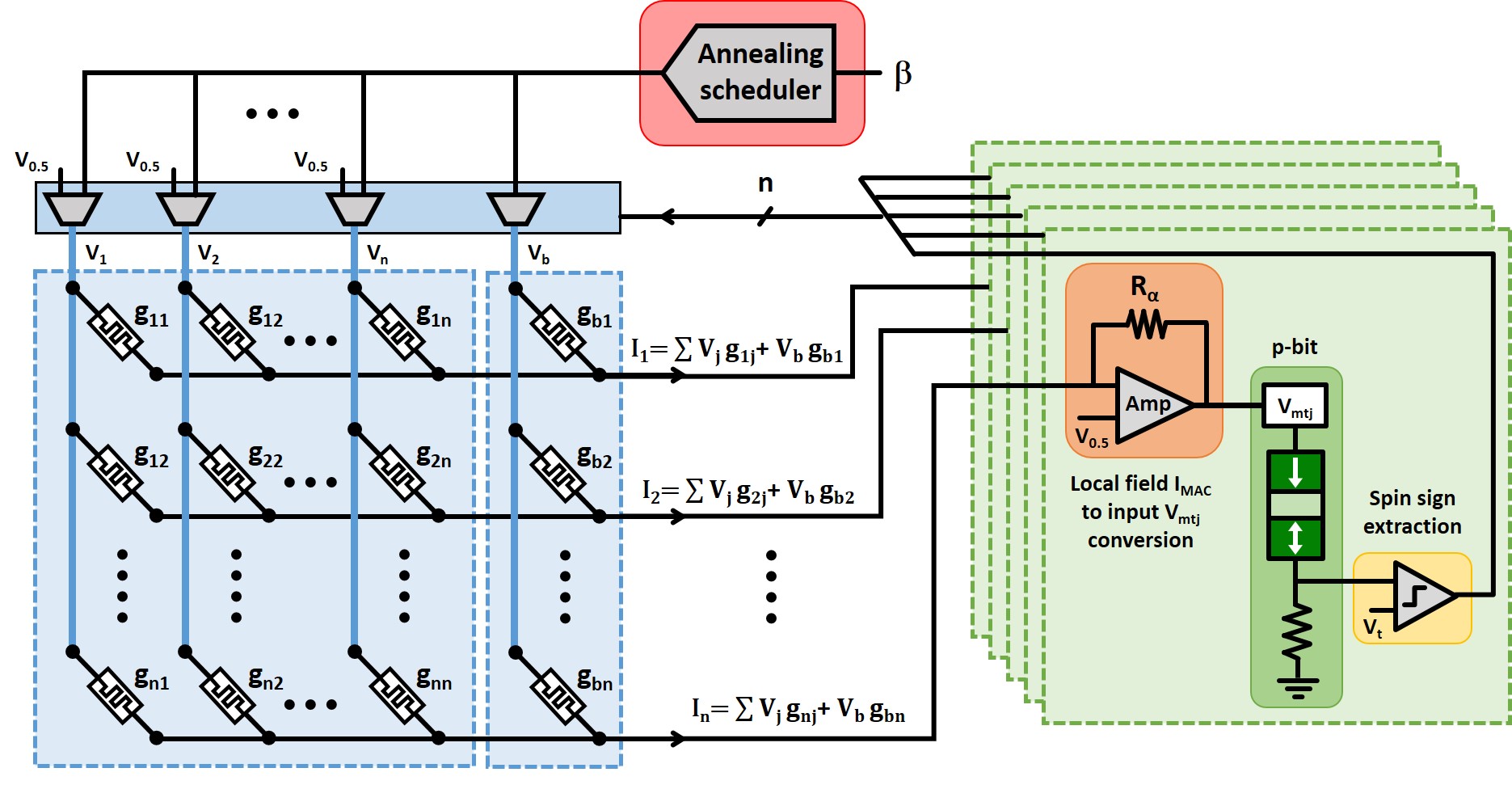}
    \caption{\textbf{Projected circuit architecture enabling parallel updates.}  
Each row of the memristor array is connected to a dedicated SMTJ‑based p‑bit (green), which embodies one Ising spin and ultimately sets the polarity applied to the corresponding column driver. The row’s multiply‑and‑accumulate current \(I_{\text{MAC}}\) is first converted to a control voltage \(V_{\text{mtj}}\) by a local transimpedance pre‑amplifier with feedback resistor \(R_\alpha\) (orange). Thermal switching of the superparamagnetic MTJ produces a stochastic output voltage, whose sign is digitised by a comparator (yellow) and fed back to the column driver of same index. Because each of the \(n\) rows carries its own p‑bit / amplifier / comparator chain, parallel updates are supported. Two global DACs, shared by the entire network, generate the \(+\!V_{\text{read}}\) and \(-\!V_{\text{read}}\) levels; an annealing scheduler modulates their amplitude (\(\beta\)) to implement simulated annealing during optimisation.}
    \label{fig:parallel_updates}
\end{figure}

\end{document}